# Molecular beam epitaxy growth of SrO buffer layers on graphite and graphene for the integration of complex oxides

Adam Ahmed[1], Hua Wen[2], Taisuke Ohta[3], Igor Pinchuk[1], Tiancong Zhu[1], Thomas Beechem[3], Roland Kawakami[1,2,*]

[1]*Department of Physics, The Ohio State University, Columbus, Ohio 43210*
[2]*Department of Physics and Astronomy, University of California, Riverside, California 92521*
[3]*Sandia National Laboratories, Albuquerque, New Mexico 87185*

**Abstract**

We report the successful growth of high-quality SrO films on highly-ordered pyrolytic graphite (HOPG) and single-layer graphene by molecular beam epitaxy. The SrO layers have (001) orientation as confirmed by x-ray diffraction (XRD) while atomic force microscopy measurements show continuous pinhole-free films having rms surface roughness of <1.5 Å. Transport measurements of exfoliated graphene after SrO deposition show a strong dependence between the Dirac point and Sr oxidation. Subsequently, the SrO is leveraged as a buffer layer for more complex oxide integration via the demonstration of (001) oriented $SrTiO_3$ grown atop a SrO/HOPG stack.

*e-mail: kawakami.15@osu.edu

Oxides are known for their multitude of electronic phases and interface chemistry. Particularly, transition metal oxides (TMO) are well known for their strong electron-electron ($e$-$e$) correlations and competing order between the Coulomb and exchange interactions in the presence of crystal field splitting. From these competing energies, it is known that oxide materials have been shown to exhibit (individually or simultaneously) a wide range of thermodynamic and/or quantum phase transitions: magnetic ordering, ferroelectricity, colossal magnetoresistance, metal-insulator transitions, and high temperature superconductivity [1,2]. In addition, the extreme sensitivity to crystalline distortions and chemical compositions offers a great deal of tunability to electronic and optical features. With the advent of pulsed laser deposition (PLD) and oxide molecular beam epitaxy (MBE), pristine thin films have been grown with atomic precision and reliability [3]. The quality of oxides and oxide heterostructures are now approaching that of semiconductors allowing oxides to be integrated into electronics for their optical, piezoelectric, or dielectric properties.

Due to these novel and exciting material phases, our goal is to integrate TMOs with 2D materials such as graphene and merge the parameter space between the 2D Dirac-like properties of graphene and strongly correlated electron physics. For spintronic applications, it would be advantageous to incorporate half metal oxides (e.g $La_xSr_{1-x}MnO_3$) [4,5] onto graphene for efficient spin injection measurements [6-8]. Furthermore, for electronic devices, high-κ dielectrics or ferroelectric oxides could allow for ultrahigh doping in graphene and improve the performance of graphene field effect transistors [9-12]. To fully capitalize upon these possibilities will require very clean interfaces, which are unlikely to be achieved using standard layer transfer methodologies [13]. Direct growth of the oxide adjacent to the graphene is therefore necessitated. However, epitaxial thin film growth on graphene has been challenging due to its weakly interacting van der Waals surface, and the growth of smooth, crystalline overlayers has been extremely limited [14,15].

In response, our strategy utilizes a smooth buffer layer for integration of complex oxides onto graphene. This approach was highly successful for integration of complex oxides onto silicon. For example, high-κ dielectric $SrTiO_3$ was first grown on Si(001) using a Sr flux to desorb $SiO_2$ from the surface [16], followed by silicide formation with Sr to facilitate the growth of SrO and, subsequently, $SrTiO_3$ [17-21]. These techniques produce $SrTiO_3$-based metal-oxide-semiconductor field effect transistors (MOSFET) with gate leakage two orders of magnitude smaller than similar devices using $SiO_2$ [22]. While direct growth onto Si(001) without buffer layer has been achieved recently [23], the use of SrO(001) buffer layers still provides a facile method for complex oxide integration.

Here, a similar approach is pursued. Specifically, we demonstrate the growth of smooth (001)-oriented SrO films on both highly-ordered pyrolytic graphite (HOPG) and single-layer graphene (SLG) to be utilized as a buffer layer for further oxide integration. Characterization by reflection high energy electron diffraction (RHEED) and x-ray diffraction (XRD) show that the growth of elemental Sr in a partial pressure of oxygen produces (001)-oriented SrO films on both HOPG and SLG. While the out-of-plane orientation is well defined, the films consist of multiple grains that have differing in-plane orientations. Atomic force microscopy (AFM) is used to determine the morphology of the films, which is strongly dependent on growth temperature and growth rate. Interestingly, the morphology is different for the HOPG and SLG substrates. For growth on HOPG at room temperature, low growth rates (3.0 Å/min) result in SrO islands that partially cover the substrate whereas higher growth rates (100 Å/min) produce smooth pinhole-free films (rms roughness of ~1.2 Å). In contrast, growth of SrO on epitaxial graphene produces smooth, pinhole-free films (rms roughness of ~1.0 Å) even at low growth rates (3.0 Å/min) at room temperature suggesting more facile SrO growth atop graphene than graphite. We directly test this by mechanically exfoliating graphene onto $SiO_2$/Si substrates and find that during the same deposition, SrO on regions of SLG are smooth while SrO on regions of thicker multilayer graphene are substantially rougher. Turning to electrical properties, conductivity measurements of SrO/graphene show Dirac-like behavior with mobility and doping characteristics that can be modified with post-growth oxidation. Finally, we demonstrate the use of SrO as a buffer layer for oxide integration by successfully growing (001)-oriented $SrTiO_3$ films on SrO/HOPG.

HOPG substrates (ZYA grade from SPI) are prepared by mechanical cleavage to expose a fresh, clean surface and loaded into a MBE chamber with a base pressure of $6.0 \times 10^{-10}$ torr. The substrate is annealed at 600 °C for 30 minutes to remove surface adsorbates. Elemental Sr is evaporated from an effusion cell where typical growth rates vary from 2 to 100 Å/min. Elemental Ti is deposited from an electron beam evaporator with a typical growth rate of ~1 Å/min. Evaporation rates are measured using a quartz crystal deposition monitor. During SrO growth, oxygen ($O_2$) partial pressures are maintained at $4 \times 10^{-7}$ torr using a sapphire-sealed leak valve. For the SrO growths on epitaxial graphene, quasi-freestanding epitaxial graphene on 6H-SiC is prepared in a similar fashion to Riedl *et al*. [24]. The epitaxial graphene is confirmed to be single layer by Raman spectroscopy and low-energy electron microscopy. Unless otherwise stated, SrO is grown to a thickness of 3 nm. The $SrTiO_3$ growth is done by co-deposition whereupon Sr and Ti are evaporated in a background environment of molecular oxygen. Surface morphology and film thickness are examined using atomic force microscopy (AFM) while crystallographic orientations are probed using a combination of RHEED and XRD measurements.

Pursuit of a high quality SrO buffer layer on HOPG began with an investigation of morphology as a function of growth temperature. Figure 1 presents both AFM and RHEED images of SrO on HOPG as a function of growth temperature for films deposited at a rate of 3 Å/min. AFM images indicate that grain size increases with temperature. The surface mobility of the Sr atoms increases with temperature leading to significant island formation (See Figures 1c-e). This is most pronounced at 540 °C where the target film thickness of 3 nm produces islands with step heights varying between ~8-20 nm. Lower temperatures, in contrast, reduce island formation leading more complete coverage (see Fig 1(b)).

RHEED images confirm island growth at higher temperatures and show streaky patterns at lower temperatures, indicating good short-range ordering and atomic-scale smoothness at lower temperatures. With respect to the former, the RHEED image for a growth temperature of 540 °C (Figure 1e) possess "spots," which is characteristic of island growth processes and is consistenti with the corresponding AFM image. With respect to the latter, the RHEED patterns of both bare HOPG and the 3 nm SrO films grown between 24 °C – 300 °C are streaky and do not change with in-plane rotation. This invariance occurs because the HOPG substrate is not single-crystalline, but rather has (0001) out-of-plane orientation and domains that have different in-plane rotational orientation. Thus, the RHEED images are a superposition of all azimuthal angles. The irregular spacing of the RHEED streaks for 3 nm SrO results from the superposition of azimuthal angles. Due to this circumstance, we cannot determine whether or not there is an in-plane expitaxial relationship between the SrO and HOPG (this is determined later through studies of SrO on single-crystalline epitaxial graphene). Nevertheless, the streaks in RHEED indicate good short-range ordering and atomic-scale smoothness (locally), which is favorable for utilizing SrO as a buffer layer for further oxide growth.

Since the RHEED patterns are similar for 24 °C – 300 °C and the AFM shows better coverage at room temperature, we focus on room temperature growth to obtain complete coverage without pinholes. Specifically, the Sr deposition rate ($\gamma_{Sr}$) is increased to counteract temperature-driven surface mobility of Sr on HOPG. Figure 2 shows a comparison of AFM and RHEED patterns for growth rates of $\gamma_{Sr}$ = 3 Å/min (Figure 2a), 30 Å/min (Figure 2b), and 100 Å/min (Figure 2c). AFM images show that the surface coverage improves and the RMS roughness decreases as the growth rate is increased. RMS roughness values of 4.5Å, 2.0 Å, and 1.2 Å are obtained for growth rates of 3 Å/min, 30 Å/min, and 100 Å/min, respectively. AFM line cuts show the presence on pinholes penetrating the entire 3 nm thickness of the SrO film for growth rates of 3 Å/min and 30 Å/min. However, pinholes are completely suppressed at the high growth rate of 100 Å/min. We further investigate the surface roughness and crystal structure using RHEED. Films grown at different rates show similar RHEED patterns, which indicate that the crystal structure and film orientation match for the three samples. However, we observe that the diffraction lines become streakier with increasing growth rate suggesting that the surface becomes smoother as the Sr rate is increased. This is consistent with the AFM data.

To determine the crystallographic orientation of the SrO film, x-ray diffraction (XRD) measurements are performed on 40 nm thick films of SrO on HOPG (grown at room temperature with $\gamma_{Sr}$ = 100 Å/min). θ-2θ XRD scans (Figure 3a) show a strong SrO(002) peak and a weaker SrO(004) peak, similar to previous studies of single-crystalline films of SrO(001) [25]. We also see two additional peaks denoted with asterisks in Figure 3a (hereby referred to as the "starred" peaks). We confirm that these starred peaks come from the substrate by cleaving the HOPG crystal several times and measuring the bare substrate. Figure 3b shows that the starred peaks are present again, with the same relative intensity ratio between the HOPG crystal peaks and the starred peaks. Thus the starred peaks are not attributable to other SrO crystal or compositional phases and we conclude that the SrO films have (001) out-of-plane orientation.

To investigate whether there is a preferred in-plane orientation of the SrO(001) unit cell relative to the graphene/graphite lattice, SrO growth atop epitaxial graphene on SiC(0001) is investigated. Epitaxial graphene possesses in-plane orientation having a $(6\sqrt{3}\times6\sqrt{3})R30°$ superstructure thereby providing in-plane order not present for cleaved HOPG [24]. Additionally, growth atop graphene can vary relative to that of HOPG [26]. To this end, epitaxial graphene sample is pre-annealed in UHV at 200°C for 30 minutes to remove moisture and organic solvents. The sample is then cooled back to room temperature for subsequent growth of SrO. RHEED images of the epitaxial graphene/SiC(0001) (Figures 4a, 4b) indicate a single-crystal structure with well-defined in-plane orientation. After growth of 3 nm SrO at room temperature at a rate of 3 Å/min, the RHEED pattern (Figure 4c) becomes similar to the ones observed on HOPG. In addition, the RHEED image remains unchanged with in-plane rotation of the sample, which indicates that the pattern is a superposition of all in-plane orientations. Thus, despite the in-plane orientation of the underlying graphene, the SrO film consists of (001)-oriented grains having differing in-plane orientations.

The surface morphology of the SrO film on epitaxial graphene is examined with AFM as well. As shown in Figure 5a, the AFM image of the bare epitaxial graphene surface shows a uniform and flat surface. The rms roughness of the epitaxial graphene surface is 0.5 Å, which is the detection limit of the measurement system. After growth of SrO on epitaxial graphene (SrO/EpGr) with $\gamma_{Sr}$ = 3 Å/min, the growth shows a uniform and homogenous surface with an rms roughness of 1.0 Å. This is much smoother than SrO films on HOPG grown at the same rate (Figure 2a) but qualitatively similar to the SrO grown on HOPG at a higher growth rate of 100

Å/min (Figure 2c). For completeness, we also grow 3 nm SrO on epitaxial graphene at $\gamma_{Sr} = 100$ Å/min and obtain similar RHEED and AFM data (not shown). Therefore, we conclude that smooth, pinhole-free SrO films can be grown on graphene by performing deposition at room temperature with Sr growth rates of 3 Å/min and above.

These results suggest that SrO films can grow smoother on thin graphene compared to thick graphene (e.g. graphite). To test this hypothesis, we mechanically exfoliate graphene flakes from ZYA grade HOPG onto $SiO_2$/Si substrates. The resulting exfoliated graphene flakes have regions of both single-layer (SLG) and multi-layer graphene (MLG). SrO growth can therefore be compared as it evolves on each region during the same deposition. Figure 6a shows an optical micrograph of the SLG flake and 20 nm MLG flake used for this study. At 20 nm thickness, the MLG flake is likely to be equivalent to bulk graphite. AFM images of the SLG surface (Figure 6b) determined the rms roughness to be 2.1 Å, which is rougher than bare HOPG (0.5 Å) and epitaxial graphene/SiC (0.5 Å). Because the SLG is flexible, the roughness reflects the morphology of the underlying $SiO_2$/Si substrate. After loading into the UHV chamber, the sample is annealed at 200 °C and cooled to room temperature. Subsequently, 3 nm SrO is deposited onto the exfoliated SLG and MLG a rate of $\gamma_{Sr} = 3$ Å/min. The AFM image of SrO on SLG (Figure 6c) reveals an rms roughness of 1.8 Å, which is slightly smoother than the bare SLG. Additionally, there are no visible pinholes. In contrast, the AFM image of SrO on the neighboring MLG flake (Figure 6d) shows a much rougher film with rms roughness of 3.4 Å. The granular structure is reminiscent of growth on HOPG at low rates (Figure 2a), although the depth between any two grains for SrO growth on MLG (shown in the line cut) is not as deep as growth on HOPG. By directly comparing the SrO growth on SLG vs. MLG under identical conditions, we confirm the earlier results that slow growth rates ($\gamma_{Sr} = 3$ Å/min) are sufficient for achieving smooth films on graphene, while high growth rates ($\gamma_{Sr}$ = 100 Å/min) are needed for achieving smooth films on graphite.

To understand the effect of the SrO overlayers on the electrical properties of graphene, we fabricate exfoliated SLG devices on $SiO_2$(300 nm)/Si substrate. Gold (Au) electrodes are patterned by electron-beam lithography and the degenerately-doped Si substrate is used as a back gate to tune the carrier density. The devices are inserted into the UHV system, annealed at 100 °C for 60 min, and SrO is deposited onto the graphene at room temperature at a growth rate of $\gamma_{Sr}$ = 3 Å/min. Electrical transport measurements are then performed *in-situ* without removing the device from the UHV chamber.

As shown in Figure 7, the four-probe resistance versus gate voltage (black curve) exhibits a Dirac point (i.e. resistance maximum) at -5 V. The field effect mobility, derived from slope of conductivity vs. gate voltage [27], is 3050 cm$^2$/Vs. After deposition of 2 ML of SrO ($\gamma_{Sr}$ = 3 Å/min), the red, open circle set of data in Figure 7 shows that the Dirac point has shifted to a negative value less than -90 V. Previous studies have reported that MBE growth of Sr in a molecular oxygen background is fully oxidized with a rate of up to 64 Å/min in a $P_{O2}$ = 8.8×10$^{-8}$ torr [28], but their study relied on the sensitivity of a quartz deposition monitor. Resistance measurements of the graphene layer are a much more sensitive probe of SrO stoichiometry, as oxygen vacancies act as charged impurity scatterers. Previous studies of metal adatom dopants on graphene show drastic shifts in the Dirac point and lower mobility with as little as 0.01 ML coverage of Ti or Fe [27,29,30]. The observed gate-dependent resistance curve indicates that the SrO is not fully oxidized and the oxygen vacancies act as charged impurity scatterers. To more fully oxidize the SrO film, the device is transferred *in-situ* to another chamber (base pressure: 1×10$^{-7}$ torr) where the sample is exposed to 1 atm of $O_2$. The first exposure is for 40 minutes, the second exposure is for 10 hours (10 h 40 min total), and the third exposure is for 16 hours (26 h 40 min total). Figure 7 shows the shift in the Dirac point voltage ($V_D$) and mobility after the first exposure (blue crosses, $V_D$ = -76 V, mobility = 430 cm$^2$/Vs), after the second exposure (teal triangles, $V_D$ = -28 V, mobility = 1350 cm$^2$/Vs), and after the third exposure (magneta squares, $V_D$ = -17 V, mobility = 1740 cm$^2$/Vs). The shift of the Dirac point closer to 0 V and the sharpening of the resistance peak indicate that charged impurities (i.e. oxygen vacancies) are being removed and the mobility is increasing. However, even with ~27 hours of post-oxidation, the mobility and the Dirac point do not fully recover to their original values. This could be due to charged impurity scattering from oxygen vacancies of other defects in the SrO film. Future studies should focus on increasing mobility by increasing the SrO film quality through reduction of oxygen vacancies and structural defects.

Finally, the use of SrO as a buffer layer to integrate SrTiO$_3$ with HOPG is demonstrated. First, a 3 nm SrO layer is grown on HOPG at room temperature with a Sr rate of 100 Å/min. For the subsequent growth of SrTiO$_3$, the Sr and Ti rates are flux matched to obtain the correct atomic ratio of Sr:Ti as 1:1. Practically, this requires that the ratio of Sr:Ti deposition rates are 3.19:1. The Sr and Ti are co-deposited with $\gamma_{Sr}$ = 3 Å/min, oxygen background pressure of 2×10$^{-7}$ torr and substrate temperature of 650°C.[31] After depositing ~25 nm of SrTiO$_3$, the crystallographic orientation of the film is probed using XRD and RHEED. The XRD scan of the SrTiO$_3$ film (see Figure 8) possesses the SrTiO$_3$(002) peak indicative of a film oriented along (001). The RHEED image of the SrTiO$_3$ film (see Figure 8 inset), meanwhile, was found to be invariant with in-plane rotation and have a streaky response, characteristics qualitatively similar to that observed in the underlying SrO buffer films. Taken together, these results are consistent with epitaxial growth of SrTiO$_3$(001) on SrO(001) with a 45° in-plane rotation of the cubic unit cell and 7% lattice mismatch between the layers ($a_{SrO}$ = 5.14 Å, $a_{SrTiO3}$ = 3.906 Å), as has been previously reported [21,32].

In summary, we have grown (001)-oriented films of SrO on HOPG and graphene. Smooth, pinhole-free films are obtained through room temperature growth with a high Sr growth rate of 100 Å/min for HOPG. For growth on graphene, in contrast, smooth pinhole-free films are obtained for a wide range of Sr growth rates (3 – 100 Å/min). These trends are verified through growth on exfoliated SLG and MLG flakes, which exhibit smooth SrO films on SLG and rough films on MLG during the same—low growth rate—deposition. While the out-of-plane (001) orientation is well-defined, the film consists of grains having differing in-plane orientations. No in-plane epitaxial relationship between the SrO and graphene lattices has been identified. The electrical properties of SrO/graphene are characterized by *in-situ* four probe measurements. Oxygen vacancies in SrO introduce charged impurity scattering which shifts the Dirac point and reduces mobility, but the scattering can be substantially reduced through post-oxidation of the SrO film. Finally, to show the utility of SrO as a buffer layer for oxide integration, 25 nm of $SrTiO_3$ was successfully grown on 3 nm SrO/HOPG resulting in $SrTiO_3$ films having (001) orientation. Thus, the growth of SrO on HOPG and graphene is a significant advance for the integration of complex oxides onto graphene.

**Acknowledgements**

This work was supported by ONR (N00014-14-1-0350) and C-SPIN STARnet, a Semiconductor Research Corporation program sponsored by MARCO and DARPA. A. A. acknowledges support from ENCOMM, and H. W. acknowledges support from NRI-NSF (DMR-1124601). The authors are appreciative of Guild Copeland for aid in synthesis of the epitaxial graphene. This work was supported by the LDRD program at Sandia National Laboratories (SNL). Sandia

**Figure Captions**

**Figure 1.** SrO growth on HOPG as a function of substrate temperature. Films are characterized by AFM (top row) and RHEED (bottom row) images. AFM scan areas are 1.9×1.9 $\mu m^2$. **(a)** Room temperature images of HOPG before deposition. The next four images are for 3 nm SrO films grown at temperatures of **(b)** 24 °C, **(c)** 150 °C, **(d)** 300 °C, and **(e)** 540 °C. Note: color scales vary between images.

**Figure 2.** SrO growth on HOPG as a function of deposition rate. All AFM images (left column) are 500×500 $nm^2$ (color bar scale: 0 to 3.5 nm) and show room temperature growth of SrO on HOPG for Sr growth rates of **(a)** 3 Å/min, **(b)** 30 Å/min, and **(c)** 100 Å/min. Only **(c)** shows complete film coverage whereas exposed areas of the HOPG substrate (step edge heights ~3 nm) are evident in **(a)** and **(b)**. AFM line cuts are shown in the middle column and RHEED images are shown in the right column. With increasing Sr rate, the qualitative features of the diffraction images become streakier indicating a flatter and more continuous surface.

**Figure 3.** XRD analysis of SrO films on HOPG. **(a)** XRD scan of 40 nm SrO film on HOPG. Only the (00$\ell$) direction, where $\ell$ is an integer, is present in the scan indicating a single out-of-plane crystalline direction. **(b)** XRD scan of the freshly cleaved ZYA grade HOPG crystal. The starred peaks are of unknown origin and are present in the HOPG substrate.

**Figure 4**. RHEED analysis of SrO films on epitaxial graphene. **(a , b)** RHEED patterns of epitaxial graphene along two in-plane directions before growth, **(c)** RHEED pattern after growth of 3 nm of SrO. This pattern does not change upon in-plane rotation of the sample.

**Figure 5.** AFM analysis of SrO films on epitaxial graphene. **(a)** Bare epitaxial graphene (EpGr) surface. The blue arrow indicates the region where the line cut was taken. The epitaxial graphene surface has an rms roughness of 0.5 Å which is the resolution limit of the AFM. **(b)** 3 nm SrO on epitaxial graphene (SrO/EpGr). The SrO grows uniformly on the epitaxial graphene, and the rms roughness of the SrO film is 1.0 Å.

**Figure 6**. Growth of SrO at room temperature ($\gamma_{Sr}$ = 3 Å/min) on exfoliated graphene on $SiO_2$/Si. **(a)** Optical image of the single-layer graphene (SLG) outlined by the black dashed line and 20 nm thick multilayer graphene (MLG). **(b)** AFM of the bare SLG flake. The rms roughness of the flake is 2.1 Å. **(c)** Growth of 3 nm SrO/SLG. The growth is uniform over the surface, where the average rms roughness of the SrO is 1.8 Å. **(d)** Growth of 3 nm SrO/MLG. The granular growth is similar to growth on HOPG (c.f. Figure 2a).

**Figure 7.** Four-probe resistance of graphene as a function of back gate voltage. The black, solid line is a scan before any SrO is deposited on top of the device. The Dirac point is located at -5V. The open red circles are for as-deposited 2 ML of SrO. The Dirac point is significantly shifted out of the measurement range. The other curves correspond to the device response after the SrO overlayer has been exposed to 1 atm of $O_2$ (blue cross hairs: 40 minutes, teal triangles: additional 10 hours, magenta squares: additional 16 hours).

**Figure 8.** XRD scan of 25 nm $SrTiO_3$ on 3 nm SrO on HOPG. The $SrTiO_3$(002) peak is observed, indicating that the film is oriented along (001). Inset: RHEED image of $SrTiO_3$ exhibiting a streaky pattern.

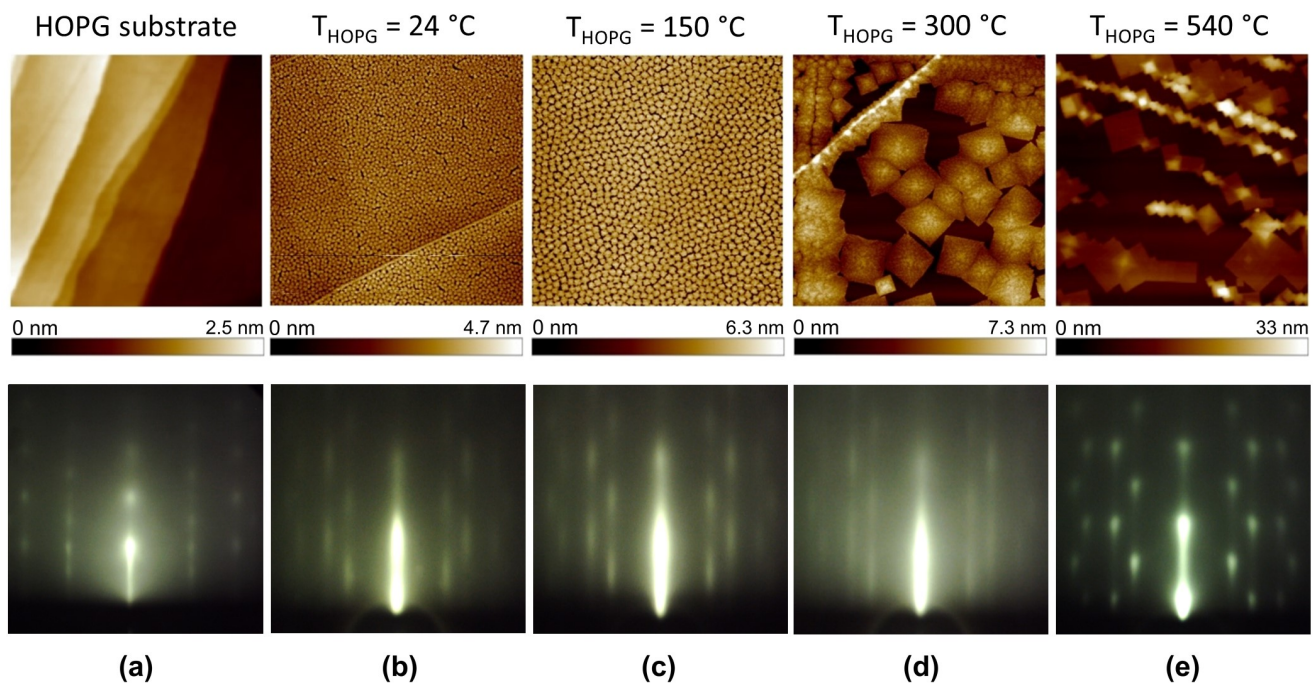

**Figure 1.** SrO growth on HOPG as a function of substrate temperature. Films are characterized by AFM (top row) and RHEED (bottom row) images. AFM scan areas are 1.9×1.9 $\mu m^2$. **(a)** Room temperature images of HOPG before deposition. The next four images are for 3 nm SrO films grown at temperatures of **(b)** 24 °C, **(c)** 150 °C, **(d)** 300 °C, and **(e)** 540 °C. Note: color scales vary between images.

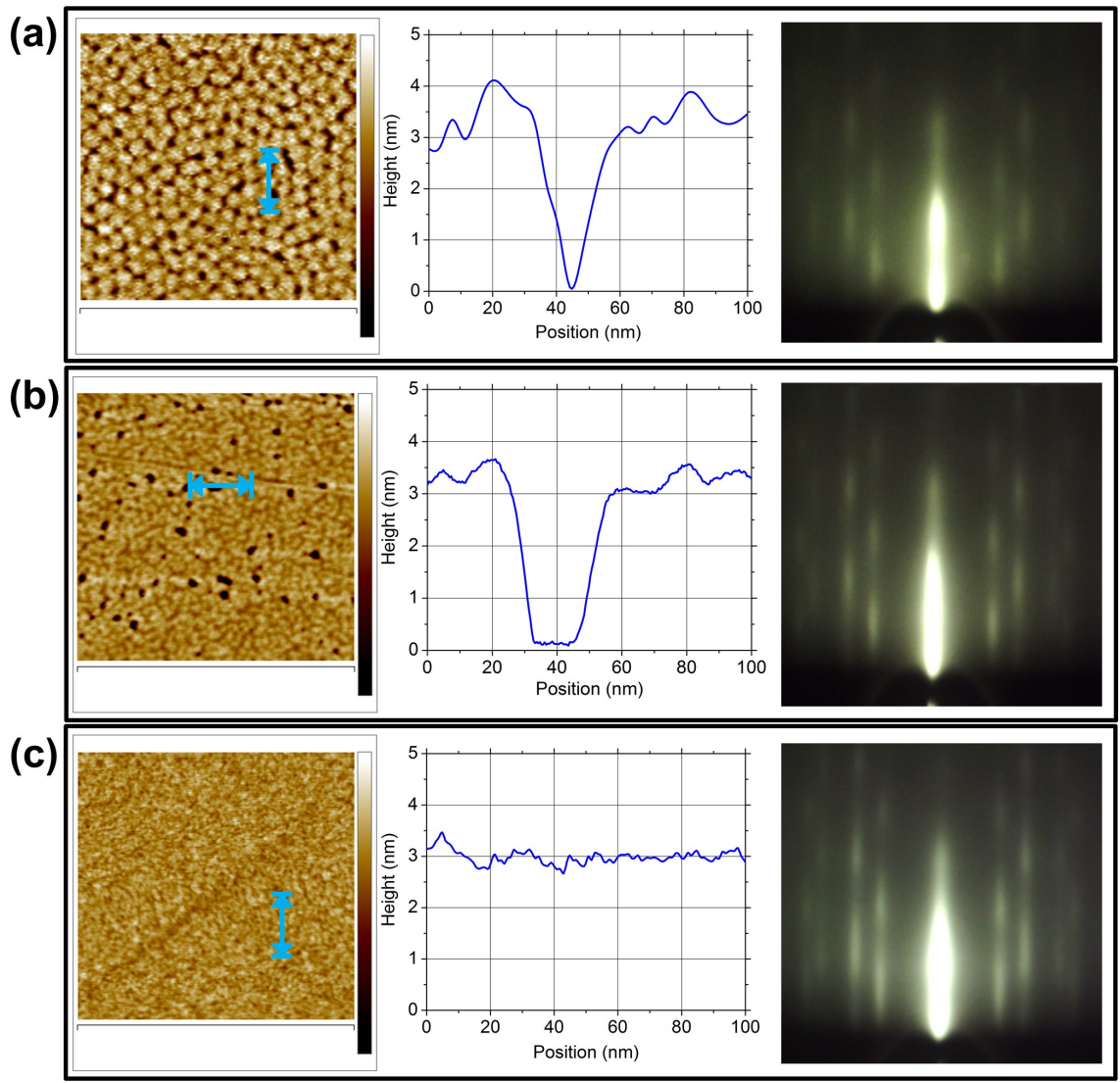

**Figure 2.** SrO growth on HOPG as a function of deposition rate. All AFM images (left column) are 500×500 nm² (color bar scale: 0 to 3.5 nm) and show room temperature growth of SrO on HOPG for Sr growth rates of **(a)** 3 Å/min, **(b)** 30 Å/min, and **(c)** 100 Å/min. Only **(c)** shows complete film coverage whereas exposed areas of the HOPG substrate (step edge heights ~3 nm) are evident in **(a)** and **(b)**. AFM line cuts are shown in the middle column and RHEED images are shown in the right column. With increasing Sr rate, the qualitative features of the diffraction images become streakier indicating a flatter and more continuous surface.

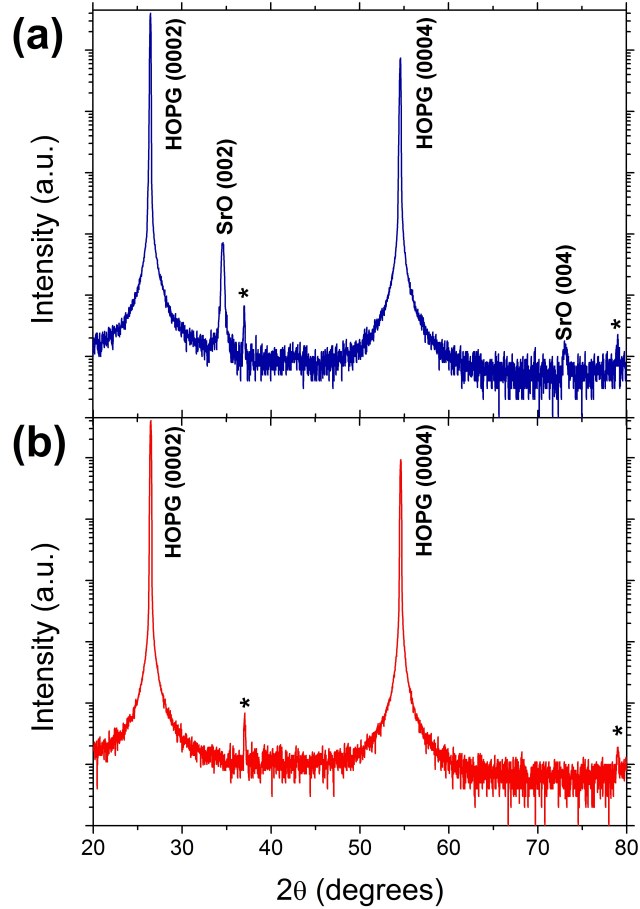

**Figure 3.** XRD analysis of SrO films on HOPG. **(a)** XRD scan of 40 nm SrO film on HOPG. Only the (00ℓ) direction, where ℓ is an integer, is present in the scan indicating a single out-of-plane crystalline direction. **(b)** XRD scan of the freshly cleaved ZYA grade HOPG crystal. The starred peaks are of unknown origin and are present in the HOPG substrate.

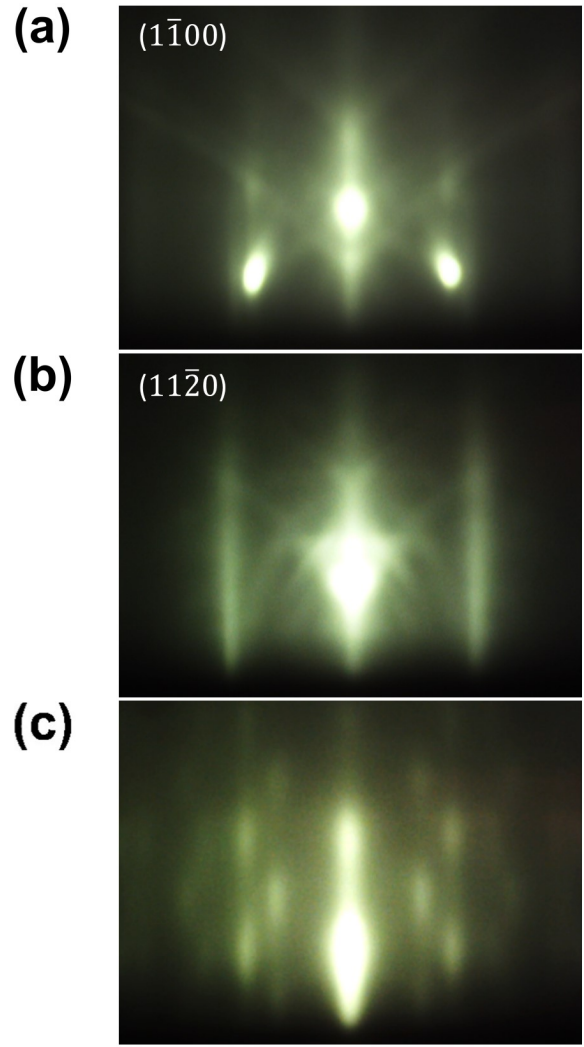

**Figure 4**. RHEED analysis of SrO films on epitaxial graphene. **(a , b)** RHEED patterns of epitaxial graphene along two in-plane directions before growth, **(c)** RHEED pattern after growth of 3 nm of SrO. This pattern does not change upon in-plane rotation of the sample.

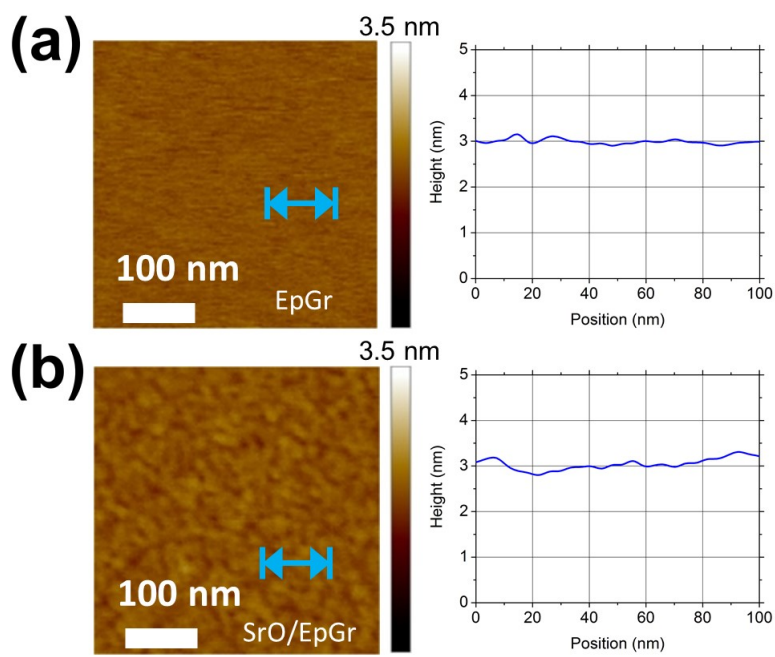

**Figure 5.** AFM analysis of SrO films on epitaxial graphene. **(a)** Bare epitaxial graphene (EpGr) surface. The blue arrow indicates the region where the line cut was taken. The epitaxial graphene surface has an rms roughness of 0.5 Å which is the resolution limit of the AFM. **(b)** 3 nm SrO on epitaxial graphene (SrO/EpGr). The SrO grows uniformly on the epitaxial graphene, and the rms roughness of the SrO film is 1.0 Å.

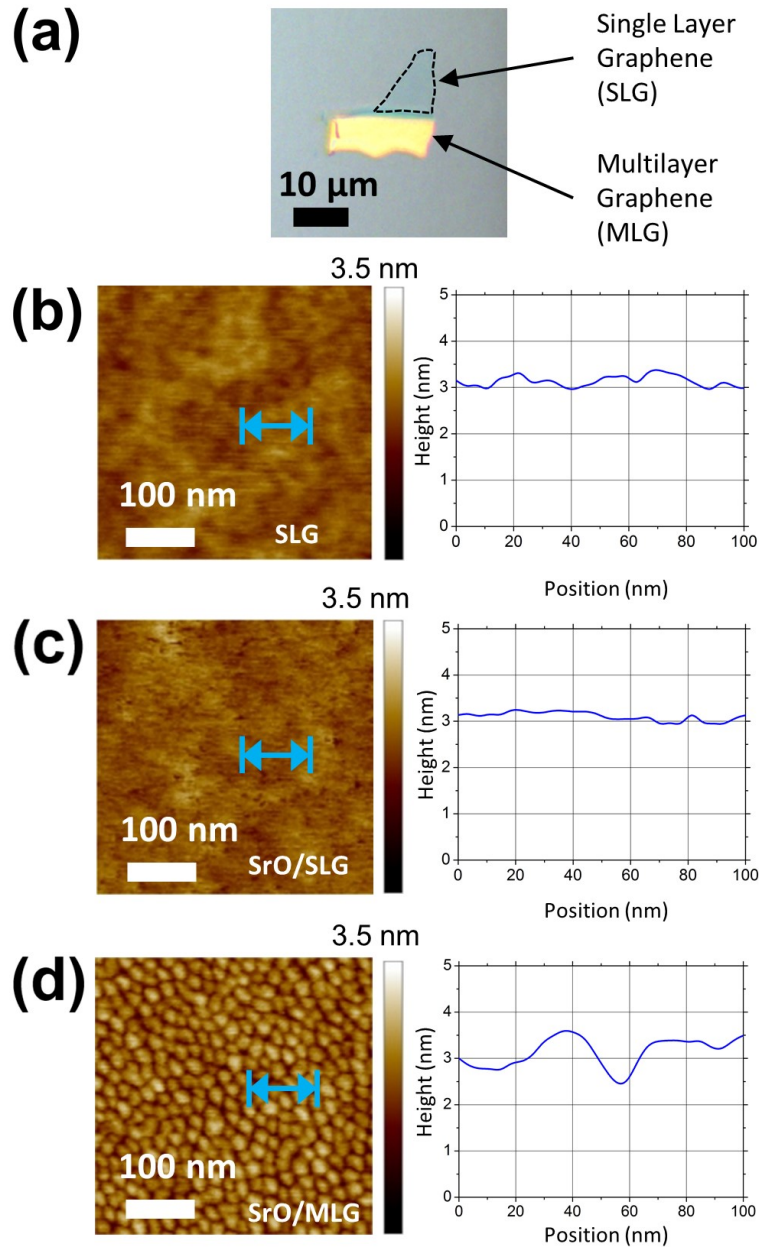

**Figure 6**. Growth of SrO at room temperature ($\gamma_{Sr}$ = 3 Å/min) on exfoliated graphene on SiO$_2$/Si. **(a)** Optical image of the single-layer graphene (SLG) outlined by the black dashed line and 20 nm thick multilayer graphene (MLG). **(b)** AFM of the bare SLG flake. The rms roughness of the flake is 2.1 Å. **(c)** Growth of 3 nm SrO/SLG. The growth is uniform over the surface, where the average rms roughness of the SrO is 1.8 Å. **(d)** Growth of 3 nm SrO/MLG. The granular growth is similar to growth on HOPG (c.f. Figure 2a).

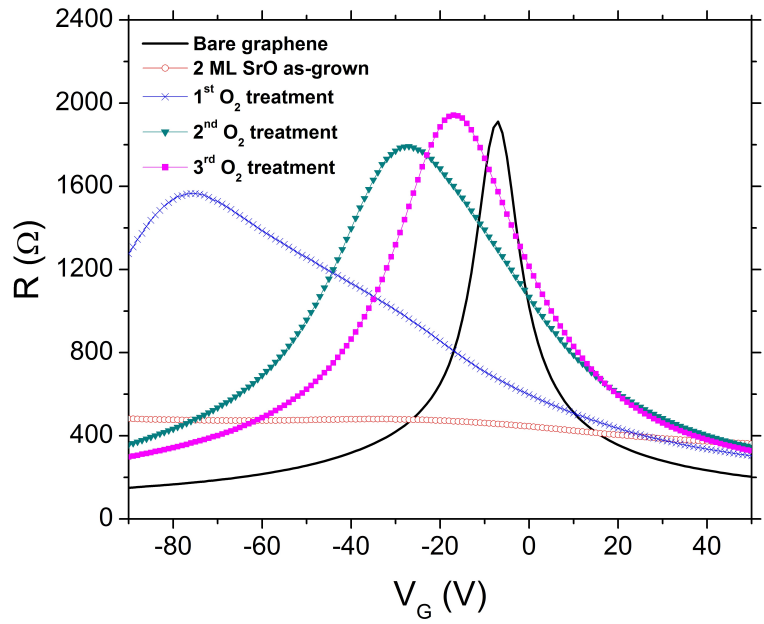

**Figure 7.** Four-probe resistance of graphene as a function of back gate voltage. The black, solid line is a scan before any SrO is deposited on top of the device. The Dirac point is located at -5V. The open red circles are for as-deposited 2 ML of SrO. The Dirac point is significantly shifted out of the measurement range. The other curves correspond to the device response after the SrO overlayer has been exposed to 1 atm of $O_2$ (blue cross hairs: 40 minutes, teal triangles: additional 10 hours, magenta squares: additional 16 hours).

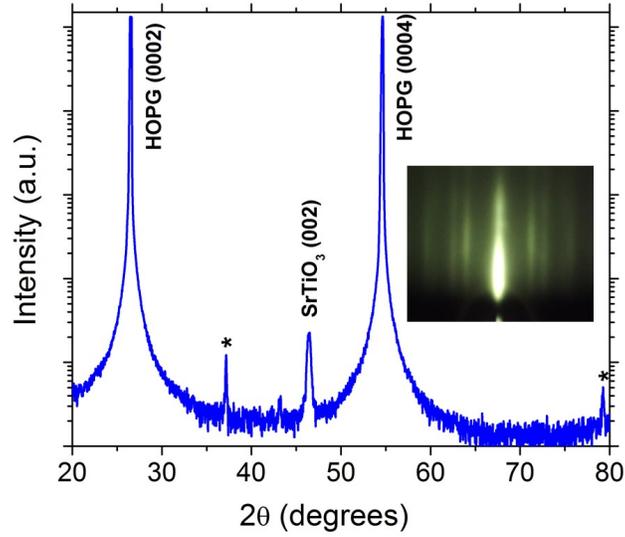

**Figure 8.** XRD scan of 25 nm SrTiO$_3$ on 3 nm SrO on HOPG. The SrTiO$_3$(002) peak is observed, indicating that the film is oriented along (001). Inset: RHEED image of SrTiO$_3$ exhibiting a streaky pattern.